\documentclass[aps,prc,twocolumn,showpacs,floatfix,nofootinbib]{revtex4} 
\usepackage{graphicx}
\usepackage{dcolumn}
\usepackage{amsmath}
\usepackage{mathptmx}
\usepackage{booktabs}
\begin{document}

\title{Evidence for the pair-breaking process in $^{116,117}$Sn}

\author{U.~Agvaanluvsan$^{1,2}$, 
A.~C. Larsen$^3$\footnote{Electronic address: a.c.larsen@fys.uio.no}, M.~Guttormsen$^3$, R.~Chankova$^{4,5}$, G.~E.~Mitchell$^{4,5}$,
A.~Schiller$^6$, S.~Siem$^3$, and A.~Voinov$^6$}

\affiliation{$^1$Stanford University, Palo Alto, California 94305 USA}
\affiliation{$^2$MonAme Scientific Research Center, Ulaanbaatar, Mongolia}
\affiliation{$^3$Department of Physics, University of Oslo, N-0316 Oslo, Norway}
\affiliation{$^4$Department of Physics, North Carolina State University, Raleigh, NC 27695, USA}
\affiliation{$^5$Triangle Universities Nuclear Laboratory, Durham, NC 27708, USA}
\affiliation{$^6$Department of Physics, Ohio University, Athens, OH 45701, USA}

\date{\today}

\begin{abstract}
The nuclear level densities of $^{116,117}$Sn below the neutron separation energy have been determined experimentally from the ($^3$He,$\alpha \gamma$) and ($^3$He,$^3$He$^\prime\gamma$) reactions, respectively. The level densities show a characteristic exponential increase and a difference in magnitude due to the odd-even effect of the nuclear systems. In addition, the level densities display pronounced step-like structures that are interpreted as signatures of subsequent breaking of nucleon pairs. 
\end{abstract}  

\maketitle

\section{Introduction}
Nuclear level densities are important for many aspects of fundamental and applied nuclear physics, including calculations of nuclear reaction cross sections. The level density of excited nuclei is an average quantity that is defined as the number of levels per unit energy. The majority of data for nuclear level densities are obtained from  two energy regions. At low excitation energy, the level density is obtained directly from counting of low-lying levels~\cite{Fir96}. As the excitation energy increases, the level density becomes large and individual levels are often not resolved in experiments. Therefore, the direct counting method becomes impossible. Nuclear resonances at or above the nucleon binding energy provide another source of level density data~\cite{Mughabghab}. Between these two excitation energy regions, the level density is often interpolated using phenomenological formulas ~\cite{Gil65,Egi88,Egidy}.  It is in this energy region the present measurements focus.  

Recently, an extension of the sequential extraction method, now referred to as the Oslo method, was developed by the Oslo Cyclotron Group. The Oslo method permits a simultaneous determination of the level density  and the radiative strength function \cite{Hen95,Sch00a}. For both of these quantities, the experimental results cover an energy region where there is little information available and data are difficult to obtain. However, the limitation of this method is that the results must be normalized to existing data -- from the discrete levels and neutron resonance spacings for the level density and to the total radiative width for the radiative strength function. Thus, the new and main achievement of the Oslo method is to establish the functional form of the level density and the radiative strength function in the above specified energy region.  

In this work we present results for the level density in $^{116,117}$Sn for the excitation energy $0<E<S_n-1$ MeV. The radiative strength functions of $^{116,117}$Sn have been published elsewhere~\cite{Sn_PRL}. The experimental set-up is briefly described in Section II, followed by a discussion of the analysis and normalization procedure. The experimental results for the level density are given in Section III, and the determination of various thermodynamic quantities are presented in Section IV. Conclusions are drawn in Section V. 

\section{Experimental procedure and data analysis} 

The experiment was carried out at the Oslo Cyclotron Laboratory (OCL) using a 38-MeV $^3$He beam. The self-supporting $^{117}$Sn target had a thickness of 1.9 mg/cm$^{2}$. The reaction channels $^{117}$Sn($^3$He,$\alpha\gamma$)$^{116}$Sn and $^{117}$Sn($^3$He,$^3$He$^\prime\gamma$)$^{117}$Sn were studied. 

The experiment ran for about 11 days with an average beam current of $\approx$ 1.5 nA. Particle-$\gamma$ coincidence events were detected using the CACTUS multidetector array. The charged particles were measured with eight Si particle telescopes placed at 45$^\circ$ with respect to the beam direction. Each telescope consists of a front Si $\Delta E$ detector with thickness 140 $\mu$m and a back Si(Li) $E$ detector with thickness 3000 $\mu$m. An array of 28 collimated NaI $\gamma$-ray detectors with a solid-angle coverage of $\approx$ 15\% of 4$\pi$ was used. In addition, one Ge detector was used in order to estimate the spin distribution and determine the selectivity of the reaction. The typical spin range is $I\sim 2-6 \hbar$.

\begin{figure}[htb]
\includegraphics[totalheight=11.5cm,angle=0]{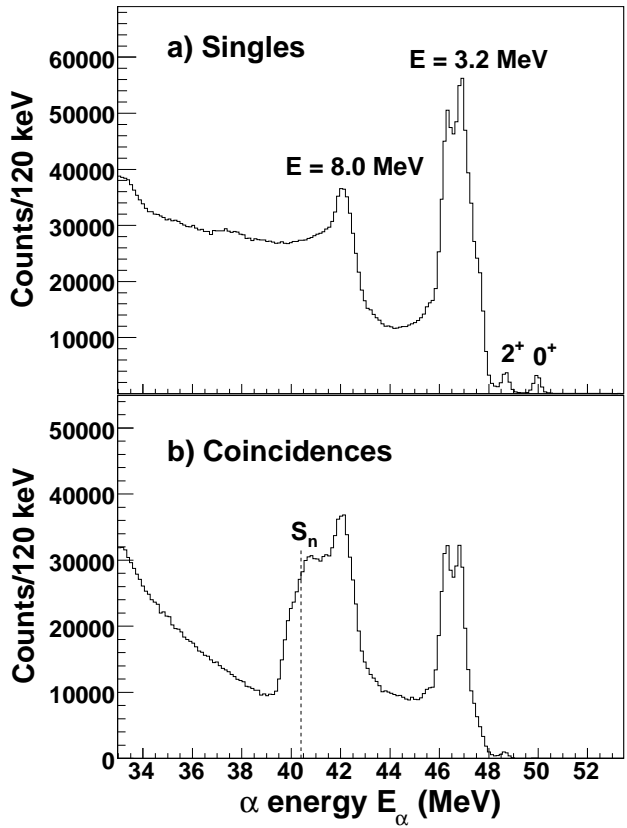}
\caption{a) Singles $\alpha$-particle spectrum, and b)
 $\alpha$-$\gamma$ coincidence spectrum from the
$^{117}$Sn($^3$He, $\alpha$)$^{116}$Sn reaction.}
\label{alphasp}
\end{figure}

Figure~\ref{alphasp} shows the singles $\alpha$-particle spectrum (upper panel) and the $\alpha$-$\gamma$ coincidence spectrum (lower panel) for $^{116}$Sn. The two peaks denoted by $0^+$ and $2^+$ are the transfer peaks to the ground state and the first excited state, respectively. The strong transfer peak at $E = 3.2$ MeV is composed of many states found to be the result of 
pick-up of high-$j$ neutrons from the g$_{7/2}$ and h$_{11/2}$ orbitals~\cite{Sch67,Yag68}. Another strong transfer peak centered around $E =8.0$ MeV is new and may indicate the neutron pick-up from the g$_{9/2}$ orbital. The counts in the coincidence spectrum decrease for excitation energies higher than $S_n$ due to lower $\gamma$-ray multiplicity when the neighboring $A-1$ isotope is populated at low excitation energy.

The particle--$\gamma$-ray coincidence spectra were unfolded using the response functions for the CACTUS detector array~\cite{Gut96}. The first-generation matrix (the primary-$\gamma$-ray matrix) contains only the first $\gamma$-rays emitted from a given excitation energy bin; this matrix is obtained  by a subtraction procedure described in Ref.~\cite{Gut87}. This procedure is justified if the  $\gamma$ decay from any excitation energy bin is independent of the method of formation -- either directly by the nuclear reaction or indirectly  by  $\gamma$ decay  from higher lying states following the initial reaction. This assumption is clearly  valid if the same states are populated via the two processes, since $\gamma$-decay branching ratios are properties of levels. When different states are populated, the assumption may not hold. However, much evidence suggests~\cite{Hen95} that statistical $\gamma$ decay is governed only by the $\gamma$-ray energy and the density of states at the final energy.

In the data analysis, the particle-$\gamma$-ray coincidence matrix is prepared and the particle energy is transformed into excitation energy using the reaction kinematics. The rows of the coincidence matrix correspond to the excitation energy in the residual nucleus, while the columns correspond to the $\gamma$-ray energy. All the $\gamma$-ray spectra for various initial excitation energies are unfolded using the known response functions of the CACTUS array~\cite{Gut96}. The first-generation $\gamma$-ray spectra, which consist of primary $\gamma$-ray transitions, are then obtained for each excitation energy bin~\cite{Gut87}. 

The first-generation $\gamma$-ray spectra for all excitation energies form a matrix $P$, hereafter referred to as the first-generation matrix. The entries of the first-generation matrix $P$ are the probabilities $P(E,E_\gamma)$ that a $\gamma$-ray of energy $E_\gamma$ is emitted from an excitation energy $E$. This matrix is the basis for the simultaneous extraction of the radiative strength function and the level density. According to the Brink-Axel hypothesis~\cite{Bri55,Axe62}, a giant electric dipole resonance can be built on every excited state with the same properties as the one built on the ground state, that is,  the radiative strength function is independent of excitation energy and thus of temperature. Many theoretical models do include a temperature dependence of the radiative strength function~\cite{Kad83,Ger98}. However, the temperature dependence is weak and the temperature change in the energy region under consideration here is rather small. Therefore, the temperature dependence is neglected in the Oslo method.

The first-generation matrix is factorized into the radiative transmission coefficient, which is dependent only on  $E_{\gamma}$, and on the level density, which is a function of the excitation energy of the final states $E-E_{\gamma}$:
\begin{equation}
\label{probab}
P(E,E_\gamma)\propto {\cal T}(E_\gamma)\rho(E-E_\gamma).
\end{equation}
The functions $\rho$ and $\cal T$ are obtained iteratively by a  globalized fitting procedure~\cite{Sch00a}. The goal of the iteration is to determine these two functions at $\sim N$ energy values each; the product of the two functions is known at $\sim N^2/2$ data points contained in the first-generation matrix. The globalized fitting to the data points determines
the functional form for $\rho$ and $\cal T$. The results must  be normalized because the entries of the matrix $P$ are invariant under the transformation~\cite{Sch00a}
\begin{eqnarray}
\tilde{\rho}(E-E_\gamma)&=&A\exp[\alpha(E-E_\gamma)]\,\rho(E-E_\gamma),
\label{eq:array1}\\
\tilde{\cal T}(E_\gamma)&=&B\exp(\alpha E_\gamma) {\cal T}(E_\gamma).
\label{eq:array2}
\end{eqnarray}

In the  final step, the transformation parameters $A$, $B$, and $\alpha$ which correspond to the most physical solution must be determined. Details of the normalization procedure are described in several papers reporting the results of the Oslo method, see for example~\cite{Sch00b,Agv04}. In the following, we will focus on the level density and thermodynamic properties.

\section{Level densities} 
\label{LD}
The coefficients $A$ and $\alpha$ relevant for the nuclear level density are determined from normalizing the level density to the low-lying discrete levels and the neutron resonance spacings just above the neutron separation energy $S_n$. For $^{117}$Sn, we used s- and p-wave resonance level spacings ($D_0$, $D_1$) taken from~\cite{Mughabghab} to calculate the total level density at $S_n$ (see Ref.~\cite{Sch00a} for more details on the calculation). Since there is no experimental information about $D_0$ or $D_1$ for $^{116}$Sn, we estimated $\rho(S_n)$ for this nucleus based on systematics for the other tin isotopes~\cite{Mughabghab,Egi88}, and assuming an uncertainty of 50\%. The level spacings and the final values for $\rho(S_n)$ are given in Table~\ref{tab:egidy}.
\begin{table*}[!htb]
\caption{Parameters used for the back-shifted Fermi gas level density and the calculation of $\rho(S_n)$.} 
\begin{tabular}{lccccccccc}
\hline
\hline
	
Nucleus    & $E_{\mathrm{pair}}$ & $C_1$ & $a$     & $D_0$ & $D_1$ &$\sigma(S_n)$  & $S_n$ & $\rho$($S_n$) & $\eta$ \\ 
           & (MeV)  		 & (MeV)  &  (MeV$^{-1}$) & (eV) & (eV)      & (MeV)  & ($10^{5}$ MeV$^{-1}$)  &        \\
\hline

$^{116}$Sn  & 2.25 & $-1.44$ & 13.13 & $-$     & $-$    & 4.96   & 9.563  & 4.12(206)  & 0.46     \\
$^{117}$Sn  & 0.99 & $-1.44$ & 13.23 & 507(60) & 155(6) & 4.44   & 6.944  & 0.86(25)   & 0.40     \\
\hline
\hline
\end{tabular}
\\
\label{tab:egidy}
\end{table*}

The experimental level density $\rho$ is determined from the nuclear ground state up to $\sim S_n-1$ MeV. Therefore, an interpolation is required between the present experimental data and $\rho$ evaluated at $S_n$. The back-shifted Fermi gas level density with the global parameterization of von Egidy {\sl et al.}~\cite{Egi88}, 
\begin{equation}
\label{rhoEq}
\rho(E)=\eta\frac{\mathrm{exp}(2\sqrt{aU})}{12\sqrt{2}a^{1/4}U^{5/4}\sigma},
\end{equation}
is employed for the interpolation\footnote{We chose to apply the old parameterization of \cite{Egi88} instead of the more recent one in \cite{Egidy} because new ($\gamma$,n) data~\cite{Hiro} showed that the slope of the radiative strength function in $^{117}$Sn \cite{Sn_PRL} became too steep using the values of \cite{Egidy}.}. The intrinsic excitation energy is given by $U=E-E_{\mathrm{pair}} - C_1$, where $C_1=-6.6A^{-0.32}$ MeV is the back-shift parameter and $A$ is the mass number. The pairing energy $E_{\rm pair}$ is based on pairing gap parameters $\Delta_{\mathrm p}$ and $\Delta_{\mathrm n}$ evaluated from even-odd mass differences \cite{Wapstra} according to~\cite{Dob01}. The level-density parameter $a$ and the spin-cutoff parameter $\sigma$ are given by $a~=~0.21A^{0.87}~{\rm MeV^{-1}}$ and $\sigma^{2} = 0.0888 A^{2/3} aT$, respectively. The nuclear temperature $T$ is described by $T = \sqrt {U/a}$ MeV. The constant $\eta$ is a parameter applied to ensure that  the  Fermi gas level density coincides with the neutron resonance data. All parameters employed for the Fermi gas level density are listed in Table~\ref{tab:egidy}.

\begin{figure}[bt]
\includegraphics[totalheight=13cm]{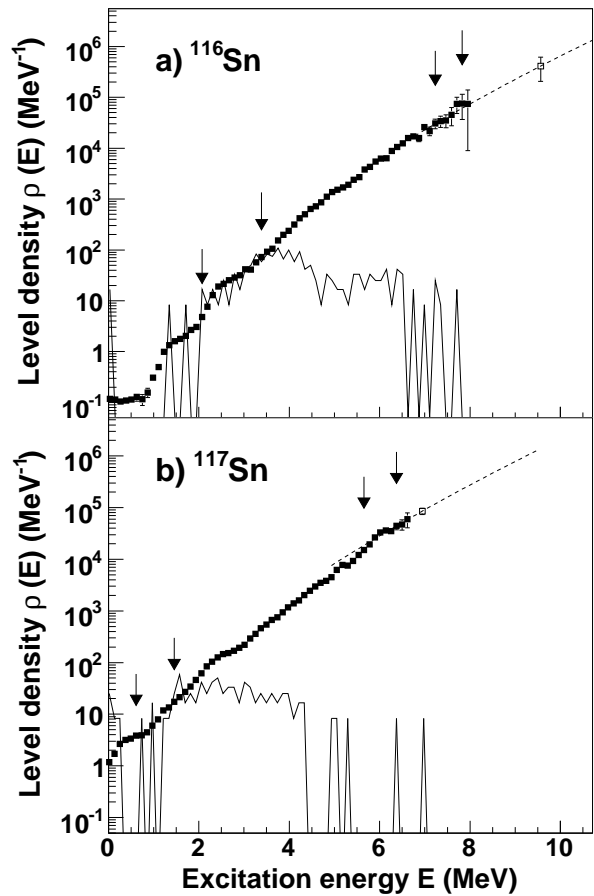}
\caption{Normalized level density of a) $^{116}$Sn and b) $^{117}$Sn. 
Results from the present work are shown as
full squares. The data points between the arrows are used for the fitting to known data. The level density at lower excitation energy obtained 
from counting of known discrete levels is shown as  a solid line, while the level density calculated from neutron resonance spacings is shown as an open square. The back-shifted Fermi gas level density used for the interpolation is displayed as a dashed line.}
\label{rhonorm}
\end{figure}

Figure \ref{rhonorm} shows the normalized level densities 
of $^{116}$Sn and $^{117}$Sn. The full squares represent
the results from the present work. The data points between the arrows are used for normalizing to 
the level density obtained from
 counting  discrete levels (solid line) and 
the level density calculated from the neutron resonance spacing 
(open square). 
The discrete level scheme is seen to be complete up to excitation energy $\approx$ 3.5 MeV 
in $^{116}$Sn beyond which the level density obtained from discrete levels starts to drop. 
For $^{117}$Sn the discrete 
level density is complete only up to $\approx$ 1.5 MeV. The new data of this work thus fill the gap between the discrete region and the calculated level density at $S_n$. 

From Fig.~\ref{rhonorm}, we observe that pronounced step-like structures are present in the level densities of both $^{116,117}$Sn. In the following section, 
nuclear thermodynamic properties are extracted using the 
present  level density results, and these structures are investigated in detail.

\section{Thermodynamic properties}

The entropy $S(E)$ is a measure of the number of ways 
to arrange a quantum system at a given excitation energy $E$. Therefore, the entropy of a nuclear system can give information on the underlying nuclear structure. The microcanonical entropy is given by 
\begin{equation}
S(E)=k_{\rm B}\ln \Omega(E),
\end{equation}
where $\Omega(E)$ is the multiplicity  of accessible states and $k_B$ is Boltzmann's constant, which we will set to unity to give a dimensionless entropy. 
 
\begin{figure}[tb]
\includegraphics[totalheight=13cm]{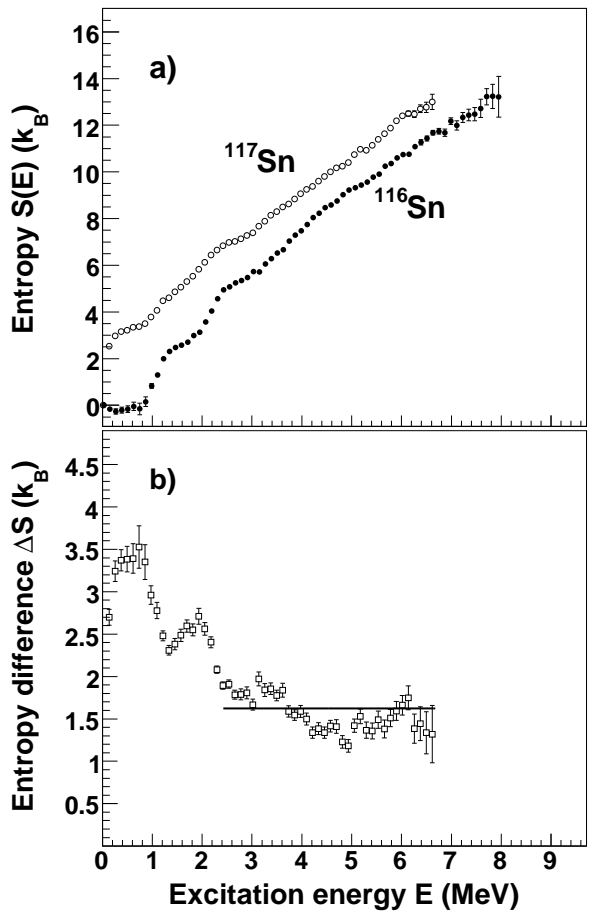}
\caption{a) Experimental entropies for $^{116,117}$Sn, and b) the entropy difference. By fitting a straight line to the entropy difference in the energy region $2.4 - 6.7$ MeV, an average value of $\overline{\Delta S} \simeq 1.6$ $k_B$ was obtained.}
\label{fig:figx}
\end{figure}

The experimental level density $\rho(E)$ is directly proportional to the multiplicity
$\Omega(E)$, which can be expressed as 
\begin{equation}
\Omega(E) = \rho(E) \cdot \left[ 2\left<J(E)\right> +1 \right].
\end{equation}
Here, $\left<J(E)\right>$ is the average spin at excitation energy $E$ and the factor $\left[ 2\left<J(E)\right> +1 \right]$ thus gives the degeneracy of magnetic substates. As the average spin is not well known at all excitation energies, we choose to omit this factor and define the multiplicity as 
\begin{equation}
\Omega(E)= \rho(E)/\rho_0, 
\end{equation}
where the denominator is determined from the
fact that the ground state of even-even nuclei 
is a well-ordered system with zero
entropy. 
The value of $\rho_0=0.135$ MeV$^{-1}$ is obtained such that	
$S =\ln \Omega \sim 0$ for the ground state region of $^{116}$Sn.  
The same $\rho_0$ is also applied for  $^{117}$Sn.
In Fig.~\ref{fig:figx} the resulting entropies of $^{116,117}$Sn are shown.

The entropy carried by the valence neutron can
be estimated by assuming that the entropy is an
extensive (additive) quantity \cite{gutt4}. With this assumption, the
experimental single neutron entropy is given by
\begin{equation}
\Delta S = S(^{117}{\rm Sn})- S(^{116}{\rm Sn}).
\end{equation}
From Fig.~\ref{fig:figx},  we observe that $\Delta S$ becomes more constant as the excitation energy increases, and above $2-3$ MeV we estimate the single neutron quasiparticle to carry
about $\Delta S \simeq 1.6$ in units of Boltzmann's constant.  This agrees 
with previous findings from the rare-earth region \cite{gutt4}. 

Both entropy curves display step-like structures superimposed on the general smooth
increasing entropy as a function of excitation energy. At these structures, the entropy increases abruptly in a small energy interval before it becomes a more slowly increasing function. 

The first low-energy bump of $^{116}$Sn is connected
to the first excited 2$^+$ state at $E=1.29$ MeV and the second excited 0$^+$ state at $E=1.76$ MeV. 
Similarly,  the first bump in the entropy of $^{117}$Sn is connected to the first excited states in this nucleus. The next structures are probable candidates for the pair-breaking process. Microscopic calculations based on the seniority model indicate that step structures in the level density can be explained by the consecutive breaking of nucleon Cooper pairs~\cite{Fe+Mo_lev}. 

The bumps present in the Sn level densities are much more outstanding than previously 
measured for other mass regions by the Oslo
group. One  explanation of the clear fingerprints could be that 
since the $Z=50$ shell is closed, the breaking of proton pairs are 
strongly hindered and thus do not smooth out the entropy  
signatures for the neutron pair breaking. Therefore, it is very likely that the structures are due to pure neutron-pair break up. 

\begin{figure}
\includegraphics[totalheight=13cm]{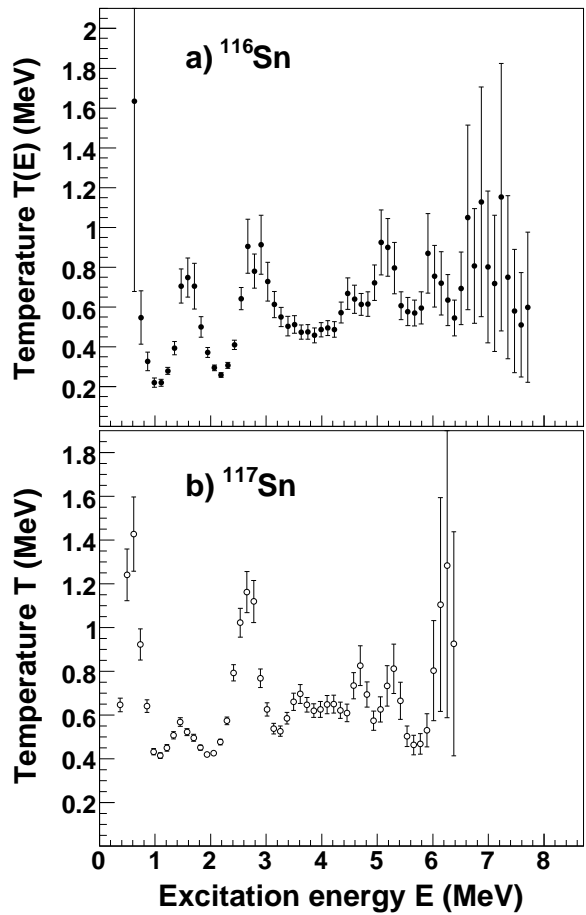}
\caption{Microcanonical temperatures of a) $^{116}$Sn, and b) $^{117}$Sn.}
\label{fig:temp}
\end{figure}

We have investigated the structures further by introducing the microcanonical temperature given by 
\begin{equation}
T = \left( \frac{\partial S}{\partial E} \right)^{-1}.
\end{equation}
By taking the derivative, small changes in the slope  of the entropy are enhanced. This is easily seen in Fig.~\ref{fig:temp}, where the microcanonical temperatures of $^{116,117}$Sn are displayed. The striking oscillations of the temperature is a thermodynamic signature for such small systems as the nucleus. Only a few quasiparticles participate in the excitation of the nucleus. Since the system is not in contact with a heat bath with a constant temperature, the system is far from the thermodynamic limit.

In Fig.~\ref{fig:temp}, the bumps below $\approx 2$ MeV are connected to the low-lying excited states in $^{116,117}$Sn. However, the peak-like structure centered around $E = 2.8$ MeV in $^{116}$Sn and around $E = 2.6$ MeV in $^{117}$Sn could be a signature of the first break-up of a neutron pair. Above $E = 4.5$ and $3.6$ MeV in $^{116,117}$Sn, respectively, the temperature appears to be constant on the average, indicating a more continuous breaking of further pairs.

Recently~\cite{gutt5}, the criticality of low-temperature transitions was investigated for rare-earth nuclei.
We apply the same method here and
investigate the probability $P$ of the system at  fixed temperature $T$ to
have  excitation energy $E$, i.e.,
\begin{equation}
P(E,T)=\Omega(E)\exp\left(-E/T\right)/Z(T),
\end{equation}
where the canonical partition function is given by
\begin{equation}
Z(T)=\int_0^\infty\Omega(E')\exp\left(-E'/T\right)\,dE'.
\end{equation}
Lee and Kosterlitz
showed \cite{LK90,LK91} that the function
$A(E,T)=-\ln P(E,T)$, for a fixed temperature $T$ in the vicinity
of a critical temperature $T_c$ of a structural transition,
will exhibit a characteristic double-minimum
structure at energies $E_1$ and $E_2$. For the critical 
temperature $T_c$, one finds $A(E_1,T_c)=A(E_2,T_c)$. 
It can be easily shown that $A$ is closely connected to the 
Helmholtz free energy, and that  this  condition is equivalent to
\begin{equation}
F_c(E_1)=F_c(E_2),
\end{equation}
which can be evaluated directly from our experimental
data. 
It should be emphasized that $F_c$ is a linearized approximation to the
Helmholtz free energy at the critical temperature $T_c$ according to
\begin{equation}
F_c(E)=E-T_cS(E).
\end{equation}

\begin{figure*}
\includegraphics[totalheight=14cm]{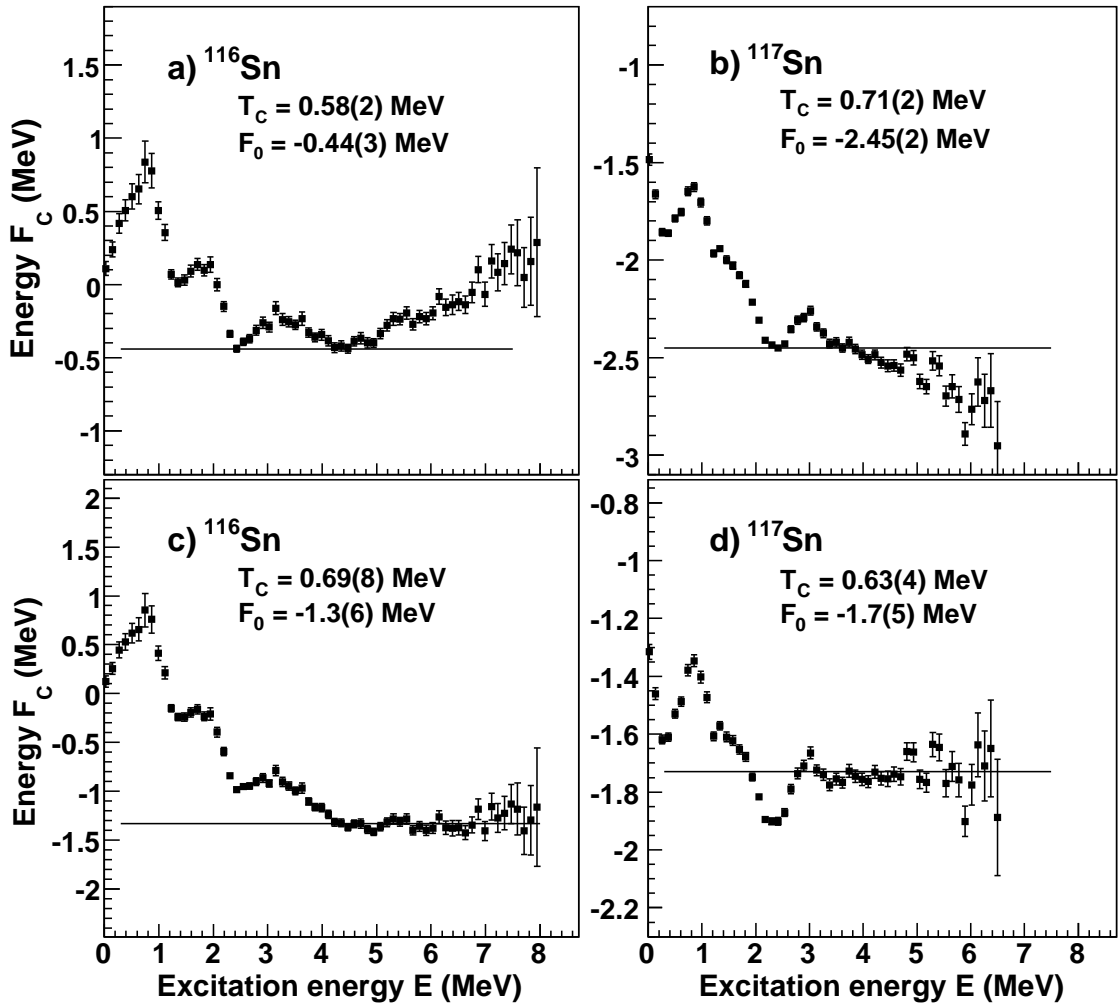}
\caption{Linearized Helmholtz free energy $F_c$ at critical temperature $T_c$ for a) $^{116}$Sn, and b) $^{117}$Sn displaying a characteristic double-minimum structure. The constant level $F_0$, which is connecting the minima, is indicated by horizontal lines. A continuous minimum of $F_c$ is shown for c) $^{116}$Sn, and d) $^{117}$Sn.}
\label{fig:figxx}
\end{figure*}

Linearized free energies $F_c$ for certain temperatures $T_c$ are 
displayed in Fig.\ \ref{fig:figxx}. 
In the upper panels, $^{116,117}$Sn  data are shown where 
the condition 
$F_c(E_1)=F_c(E_2)=F_0$ is fulfilled. Each nucleus shows a double-minimum structure, 
which we interpret as the critical temperatures 
for breaking one neutron pair.
The values found are $T_c=0.58(2)$ and $0.71(2)$ MeV for $^{116,117}$Sn,
respectively. Furthermore, we observe a potential barrier $\Delta F_c$ of about $0.25$ MeV between the two minima $E_1$ and $E_2$. The potential barrier indicates the free energy needed to go from one phase (no pairs broken) to another (one broken pair) at the constant, critical temperature $T_c$.

In the process of breaking additional pairs, the structures 
are expected to be less pronounced. 
Indeed, in the lowest panels of Fig.~\ref{fig:figxx}, the free energy 
is rather constant for excitation energies above $E\approx5.0$ and $4.2$ MeV 
for $^{116,117}$Sn, respectively. Instead of a double-minimum structure, 
a continuous minimum of $F_c$ appears for 
several MeV of excitation energies. This demonstrates
clearly that the depairing process in tin cannot be
interpreted as an abrupt structural change typical of  a first
order phase transition. For this process we evaluate the critical temperature
by a least $\chi^2$ fit of $F_0$ to the experimental data.
The excitation energies ($E_1$ and $E_2$) and temperatures 
for the phase transitions are summarized in Table~\ref{tab:tab1}.
\begin{table}[htdp]
\caption{Phase transition values deduced for $^{116,117}$Sn.}
\begin{center}
\begin{tabular}{ccccc}
\hline
\hline
 & \multicolumn{2}{c}{$^{116}$Sn} & \multicolumn{2}{c}{$^{117}$Sn}\\
Breaking of   & $E_1-E_2$& $T_c$  & $E_1-E_2$ & $T_c$ \\ 
		&	(MeV) & (MeV)	& (MeV) & (MeV) \\
\hline
one pair      & $2.4-4.2$   & 0.58(2)     & $2.3-3.7$   & 0.71(2)   \\
two or more pairs     & $5.0-8.0$   & 0.69(8)     & $4.2-6.0$   & 0.63(4)   \\
\hline
\hline
\end{tabular}
\end{center}
\label{tab:tab1}
\end{table}

According to the linearized free-energy calculations, the first neutron pair is broken for excitation energies between $2.4-4.2$ MeV in $^{116}$Sn. Comparing with the upper panel of Fig.~\ref{fig:temp}, we see that this coincides with the excitation-energy region where a peak structure is found. This bump is thought to represent the first neutron pair break-up as discussed previously in the text. The average temperature of the bump is $0.59(2)$ MeV, in excellent agreement with the deduced critical temperature $T_c = 0.58(2)$. Similarly, for $^{117}$Sn in the lower panel of Fig.~\ref{fig:temp}, we find that a peak-like structure with an average temperature of $0.74(2)$ MeV is present between $E=2.3-3.7$ MeV. Compared to the critical temperature $T_c = 0.71(2)$ MeV, the values agree satisfactory.

For the continuous break-up process characterized by a zero potential barrier ($\Delta F_c \approx 0$ MeV), we estimate for $^{116}$Sn an average microcanonical temperature of $0.75(6)$ MeV in the excitation-energy region $E=5.0 - 8.0$ MeV, in reasonable agreement with $T_c = 0.69(8)$ calculated from the linearized Helmholtz free energy. In the case of $^{117}$Sn, the average microcanonical temperature for excitation energies between $4.2-6.0$ MeV is found to be $0.64(3)$ MeV, which agrees very well with $T_c = 0.63(4)$ MeV. This gives further confidence in our interpretation of the data as two independent methods give very similar results.

\section{Conclusions}  

The nuclear level densities for $^{116,117}$Sn 
are extracted from particle-$\gamma$ coincidence measurements. 
New experimental results are reported 
for the level density in $^{116,117}$Sn
for excitation energies above 1.5 MeV and 3.5 MeV up to $S_n - 1$ MeV.  
The level densities for both nuclei display  prominent step-like  structures. The structures have been further investigated by means of thermodynamic considerations: microcanonical entropy and temperature, and through calculations of the linearized Helmholtz free energy. Both methods give consistent results, in strong favor of the pair-breaking process as an explanation of the structures.

\acknowledgments 

This research was sponsored by the National Nuclear Security Administration
under the Stewardship Science Academic Alliances program through DOE
Research Grant No. DE-FG52-06NA26194. 
U.~A. and G.~E.~M. also acknowledge support from U.S. Department of Energy Grant No. DE-FG02-97-ER41042. 
Financial support from the Norwegian Research Council (NFR) 
is gratefully acknowledged.

\end{document}